\documentclass[journal]{IEEEtran}

\ifCLASSINFOpdf
\else
   \usepackage[dvips]{graphicx}
\fi
\usepackage{url}

\usepackage[utf8]{inputenc} 
\usepackage[numbers,compress]{natbib}
\usepackage[cmex10]{amsmath}
\usepackage{amsfonts,mathrsfs,mathdots,mathtools,amsthm,amssymb,centernot}
\usepackage{caption}
\usepackage{subcaption}
\usepackage{textcomp}
\usepackage[dvipsnames]{xcolor}
\usepackage{tikz}
\usepackage{siunitx}
\usepackage{comment}
\usepackage{pgfplotstable}
\usepackage[ruled]{algorithm2e}
\usepackage{color,soul}
\usepackage{bm}	
\usepackage[inline]{enumitem}
\usepackage{nicefrac}
\usepackage{url}
\usepackage{ifthen}
\usepackage{balance}
\usepackage{xfrac}
\usepackage{dirtytalk}
\usepackage{hyperref}  
\hypersetup{colorlinks,
            linkcolor=blue,
            citecolor=blue,
            urlcolor=blue,
            anchorcolor=blue,
            filecolor=blue,
            linktocpage,
            plainpages=false,
            breaklinks=true}

\DeclarePairedDelimiter\abs{\lvert}{\rvert}%
\DeclarePairedDelimiter\norm{\lVert}{\rVert}%

\theoremstyle{definition}
\newtheorem{definition}{Definition}

\newtheorem{theorem}[definition]{Theorem}

\newcommand{\bE}{\ensuremath{\mathbb{E}}}

\newcommand{\bR}{\ensuremath{\mathbb{R}}}

\newcommand{\cD}{\ensuremath{\mathcal{D}}}

\newcommand{\cQ}{\ensuremath{\mathcal{Q}}}

\newcommand{\cW}{\ensuremath{\mathcal{W}}}
\newcommand{\cX}{\ensuremath{\mathcal{X}}}
\newcommand{\cY}{\ensuremath{\mathcal{Y}}}

\mathtoolsset{showonlyrefs}
\allowdisplaybreaks

\newcounter{relctr} 
\everydisplay\expandafter{\the\everydisplay\setcounter{relctr}{0}} 

\newcommand\labelrel[2]{%
  \begingroup
    \refstepcounter{relctr}%
    \stackrel{\textnormal{(\alph{relctr})}}{\mathstrut{#1}}%
    \originallabel{#2}%
  \endgroup
}
\AtBeginDocument{\let\originallabel\label} 

\definecolor{sara_blue}{RGB}{0, 100, 0}
\newcommand{\modified}[1]{{\color{black} #1}}

\begin{document}

\title{A Tight Context-aware Privacy Bound for Histogram Publication}

\author{Sara~Saeidian,~\IEEEmembership{Member,~IEEE,}
        Ata~Yavuzyılmaz,
        Leonhard Grosse,~\IEEEmembership{Member,~IEEE,}
        Georg~Schuppe,
        Tobias~J.~Oechtering,~\IEEEmembership{Senior~Member,~IEEE}
\thanks{This work was supported by the Swedish Research Council (VR) under grants 2023-04787 and 2024-06615, and the Digital Futures center within the collaborative project DataLEASH in Action.}
\thanks{S. Saeidian, L. Grosse, and T.~J. Oechtering are with the Division of Information Science and Engineering (ISE), School of Electrical Engineering and Computer Science, KTH Royal Institute of Technology, 100 44 Stockholm, Sweden (email: \{saeidian, lgrosse, oech\}@kth.se). S. Saeidian is also affiliated with Inria Saclay, 91120 Palaiseau, France. A. Yavuzyılmaz conducted his master’s thesis at ISE, during which a preliminary version of this work was developed~\cite{Yavuzyilmaz1911803} (email: atay@kth.se). G.~Schuppe is with SEBx, SEB (email: georg.schuppe@seb.se).}}

\maketitle

\begin{abstract}
We analyze the privacy guarantees of the Laplace mechanism releasing the histogram of a dataset through the lens of \emph{pointwise maximal leakage} (PML). While \emph{differential privacy} is commonly used to quantify the privacy loss, it is a context-free definition that does not depend on the data distribution. In contrast, PML enables a more refined analysis by incorporating assumptions about the data distribution. We show that when the probability of each histogram bin is bounded \modified{away from zero}, stronger privacy protection can be achieved for a fixed level of noise. Our results demonstrate the advantage of context-aware privacy measures and show that incorporating assumptions about the data can improve privacy-utility tradeoffs.
\end{abstract}

\begin{IEEEkeywords}
Privacy, pointwise maximal leakage, differential privacy, Laplace mechanism, histogram query. 
\end{IEEEkeywords}

\IEEEpeerreviewmaketitle

\section{Introduction}
As data-driven services continue to proliferate, the need for responsible data collection and utilization has become increasingly crucial. In response, many modern applications use mathematically rigorous frameworks to provide \emph{privacy} guarantees to their users. Among the various privacy frameworks, \emph{differential privacy} (DP)~\cite{dworkCalibratingNoiseSensitivity,dwork2014algorithmic} is the most well-studied and widely adopted. In the most basic setup, DP assumes that users' data is stored centrally and the goal is to release aggregate statistics about the data while safeguarding the privacy of all participating individuals. Over the years, this fundamental setup has been significantly expanded, leading to more sophisticated applications such as DP-enabled deep learning~\cite{abadi2016deep} and federated learning~\cite{wei2020federated}.

Despite DP's many successes, certain aspects of it can still be improved. Notably, DP is a \emph{context-free} framework, i.e., it defines privacy as a property of the data release process (called the \emph{privacy mechanism}) alone without considering the distribution of the private data. This raises important questions: Is DP universally suitable for all data distributions? Moreover, when (partial) information about the data distribution is available, can we exploit this knowledge to design mechanisms with better privacy-utility tradeoffs? 

\emph{Pointwise maximal leakage} (PML) is a recent notion of privacy that addresses these considerations and has several notable advantages. First, PML has a strong operational meaning since it is obtained by quantifying privacy risks in highly general and robust threat models~\cite{saeidian2023pointwise_it,saeidian2023pointwise_isit}. Second, PML is a \emph{context-aware} definition, that is, it depends on the data-generating distribution in addition to the privacy mechanism. For this reason, privacy mechanisms designed to satisfy PML privacy for specific data distributions often achieve greater utility than those designed to satisfy context-free definitions such as DP~\cite{grosse2024extremal}. Importantly, if no information about the data distribution is available, PML can still be applied by assessing privacy under the worst-case distribution.

The underlying philosophy of PML differs fundamentally from that of DP. Specifically, DP was designed to protect participants in a database, but no protection is guaranteed for individuals whose data is correlated with the database in other ways~\cite{dworkDifficultiesDisclosurePrevention2010a}. In contrast, PML protects features of the data with large \emph{entropy} but allows disclosing features with small entropy~\cite{inferential_privacy}.\footnote{It is well-known that all meaningful privacy definitions must allow some information disclosure to provide utility~\cite{dworkDifficultiesDisclosurePrevention2010a}.
} Despite their differences, connections between the two frameworks have been established. It has been shown that DP is equivalent to restricting the PML of every record in all databases with independent entries~\cite{inferential_privacy}. Conversely, a mechanism satisfying DP applied to a dataset with correlated entries can be trivially non-private in the sense of PML~\cite{apf24}.

This letter is part of a series of works exploring the connections between PML and DP while highlighting their differences. It extends a previous study where we analyzed a core DP mechanism, namely, the \emph{Laplace mechanism}~\cite{dworkCalibratingNoiseSensitivity} applied to the \emph{counting query}\footnote{A counting query computes the number of entries in a database that satisfy a certain predicate.} in the PML framework~\cite{inferential_privacy}. There, we showed that for the same amount of noise, databases with higher entropy achieve stronger privacy guarantees. In this work, we extend the study of the Laplace mechanism to \emph{histogram queries}~\cite{dworkCalibratingNoiseSensitivity}. Histograms serve as an important case study for various reasons. For example, in privacy-preserving statistics, histograms are often the basis for more complex algorithms, such as DP-based estimation of statistical properties over unbounded domains~\cite{karwa2018finite} and 
synthetic data generation~\cite{zhang2017privbayes, mckenna2019graphical, zhang2021privsyn}. 

\textbf{Contributions.} We compute the PML guarantees of the mechanism that releases a histogram perturbed with Laplace noise \modified{and complement our analysis with empirical comparisons against standard DP mechanisms.} From the theoretical perspective, private histograms are particularly interesting because their DP guarantees can be derived in two distinct ways. A naive approach treats a histogram with $k$ bins as $k$ independent counts, leading to a privacy parameter that scales linearly with $k$ via the basic composition theorem. A more refined analysis instead observes that modifying a single database entry affects at most two bins. Then, by using the global sensitivity, one obtains a DP bound that is independent of $k$~\cite{dwork2014algorithmic}. Interestingly, a similar phenomenon occurs with PML, even though sensitivity is \emph{not} a core concept in the PML framework. While applying the composition theorem of PML yields a leakage bound that grows with $k$, our results show that the bound can, in fact, be made independent of $k$.

\modified{Our analyses further show that: (a) histograms with more \emph{balanced} probabilities admit stronger PML guarantees for the same amount of noise; and (b) PML-privatized histograms consistently achieve higher utility than DP-privatized histograms, especially in the high-privacy regime.}

\textbf{Other related works.} 
Adding Laplace noise directly to the histogram bins works well when the number of bins $k$ is small. However, several studies have pointed out that when $k$ is large, the released histogram can become very noisy and uninformative. To address this, various methods have been developed to reduce the required noise. Broadly speaking, these approaches fall into two categories. The first category uses \emph{clustering} methods to group bins with similar counts together~\cite{xu2013differentially,acs2012differentially,kellaris2013practical,zhang2014towards}. \modified{The second category transforms the histogram or the underlying distribution into an alternative functional representation, e.g., via the Fourier transform~\cite{barak2007privacy}, wavelet transform~\cite{xiao2010differential}, Efron–Stein decomposition~\cite{yang2012differential}, or polynomial basis expansion~\cite{tao2025differentially}.} Here, we focus on the standard case of adding noise directly to the histogram since our goal is to analyze the impact of the data distribution on the privacy guarantees.  

\section{Preliminaries}
\label{sec:prelem}
\subsection{Notation}
We use uppercase letters to denote random variables, lowercase letters to denote their realizations, and calligraphic letters to denote sets. \modified{We let $X \in \cX$  represent some sensitive information, typically a dataset of individuals' records, with distribution $P_X$.\footnote{With a slight abuse of notation, we treat \modified{(conditional)} probability distributions and \modified{(conditional)}  densities interchangeably. \modified{Densities are probability mass functions when the underlying space is finite.}} We let $Y \in \cY$ denote some information released about $X$, for example, the output of a query. $Y$ is induced by a \emph{privacy mechanism} \( M_{Y \mid X} \), which is a conditional probability distribution. We also use the notation \( [n] \coloneqq \{1, \ldots, n\} \), and write \( \log(\cdot) \) for the natural logarithm.}

\subsection{Pointwise Maximal Leakage}
Pointwise maximal leakage (PML) is a recent notion of information leakage derived by analyzing privacy risks in two robust and highly general threat models: the \emph{randomized function} model~\cite{issaOperationalApproachInformation2020} and the \emph{gain function} model~\cite{alvim2012measuring,alvim2014additive}. \modified{PML is a well-behaved privacy measure that satisfies useful inequalities, such as pre-processing, post-processing, and composition inequalities~\cite{saeidian2023pointwise_it}.} 

Below, we define PML using the gain function model. For details about the randomized function model, see~\cite{issaOperationalApproachInformation2020,saeidian2023pointwise_it}. 
\begin{definition}[{\cite[Def. 3]{saeidian2023pointwise_isit}}]
\label{def:gain_function_view}
Let $Y$ be a random variable induced by the mechanism $M_{Y \mid X}$ with input $X$. The pointwise maximal leakage from $X$ to $y \in \cY$ is defined as\footnote{Both here and in Definition~\ref{def:dp}, $y$ refers to any type of output of the mechanism, for example, a scalar or a tuple. Later, we use $y^k$ to explicitly indicate that the mechanism's outcome is a tuple with $k$ components.} 
\begin{equation}
\label{eq:g-leakage}
    \ell_M(X\to y) \coloneqq \log \; \sup_{\cW, g} \; \frac{\sup_{K_{W \mid Y}} \mathbb{E} \left[g(X,W) \mid Y=y \right]}{\sup_{w' \in \cW} \bE\left[g(X, w')\right]},
\end{equation}
where the supremum is over all measurable spaces $\cW$ and non-negative measurable functions $g: \cX \times \cW \to \bR_+$ with $\sup_{w' \in \cW} \bE [g(X,w')] < \infty$, and $K_{W \mid Y}$ is the conditional distribution of $W$ given $Y$.
\end{definition}

\modified{Definition~\ref{def:gain_function_view} captures the worst-case posterior-to-prior ratio of an adversary’s expected gain, maximized over all nonnegative gain functions $g$. Note that, in general, $X$ can represent any type of sensitive data, such as an entire database or information about a single individual. However, in Section~\ref{sec:histo}, we assume that $X$ is a database.}

In~\cite{saeidian2023pointwise_isit} we showed that Definition~\ref{def:gain_function_view} simplifies to
\begin{align}
\label{eq:pml_simple}
    \ell_M(X \to y) = \log \; \sup_{x \in \cX} \; \frac{M_{Y \mid X}(y \mid x)}{M_Y(y)},
\end{align}
where $M_Y$ is the marginal of $Y$ induced by $M_{Y \mid X}$ and $P_X$.  

\subsection{Differential Privacy}
\modified{Differential privacy (DP) protects each individual's privacy by requiring that query responses returned by databases differing in a single entry (i.e., \say{neighboring} databases) have similar probabilities. Over the years, this core idea has spawned many different variants such as approximate DP~\cite{dworkOurDataOurselves2006a}, concentrated DP~\cite{bun2016concentrated,dwork2016concentrated}, and Rényi DP~\cite{mironov2017renyi}. Here, we focus on the original variant, also known as \say{pure} DP.}

Let $X = (D_1, \ldots, D_n) \in \cD^n$ be a database with $n$ entries,\footnote{\modified{We assume that the number of records $n$ is publicly known, as is standard in the \say{bounded} differential privacy model (see \cite{kiferNoFreeLunch2011}).}} where $D_i \in \cD$ denotes its $i$-th entry. Suppose databases are sampled from $\mathcal{D}^n$ according to $P_X = P_{D_1, \ldots, D_n}$. Given $x,x' \in \cD^n$, we write $x \sim x'$ to imply that $x$ and $x'$ are neighboring databases. 

\begin{definition}[Pure DP~\cite{dworkCalibratingNoiseSensitivity}]
\label{def:dp}
Given $\varepsilon \geq 0$, the mechanism $M_{Y \mid X}$ satisfies $\varepsilon$-differential privacy if
\begin{equation*}
    \sup_{y \in \cY} \; \sup_{\substack{x,x' \in \cD^n :\\ x\sim x'}} \; \log \; \frac{M_{Y \mid X}(y \mid x)}{M_{Y \mid X}(y \mid x')} \leq \varepsilon.
\end{equation*}
\end{definition}

Observe that differential privacy is a property of the mechanism $M_{Y \mid X}$ alone and does not depend on $P_X$. 

Next, we define the Laplace mechanism, which is one of the most fundamental DP mechanisms. Let $\mathrm{Lap}(b)$ denote the Laplace distribution with mean zero and scale parameter $b >0$, i.e., variance $2b^2$. 

\begin{definition}[Laplace mechanism~\cite{dworkCalibratingNoiseSensitivity}]
\label{def:laplace_mech}
Let $f : \cD^n \to \bR^k$ be a query with $\ell_1$-sensitivity 
\begin{equation*}
    \Delta_1(f) \coloneqq \sup_{x, x' \in \cD^n : x \sim x'} \norm{f(x) - f(x')}_1.
\end{equation*}
Suppose the elements of $N^k = (N_1, \ldots, N_k)$ are drawn i.i.d from $\mathrm{Lap}\left(b\right)$ with $b>0$. Then, the Laplace mechanism 
\begin{equation*}
    Y^k = f(x) + N^k, \quad x \in \cD^n, 
\end{equation*}
satisfies $\frac{\Delta_1(f)}{b}$-DP, where $Y^k = (Y_1, \ldots, Y_k)$ is the output of the mechanism.
\end{definition}

\section{PML Analysis of the Histogram Query}
\label{sec:histo}
\subsection{Theoretical Analysis}
\modified{Our first goal is to derive a tight upper bound on the leakage incurred by the Laplace mechanism.} Let $X \in \cX = \cD^n$ be a database with $n$ independent records, and suppose we want to compute its histogram with $k$ bins.\footnote{\modified{The number of bins $k$ is assumed to be publicly known.}} Let $ \{h_j\}_{j=1}^k $ be a collection of indicator functions, where $ h_j : \mathcal{D} \to \{0, 1\} $ determines whether or not a data point belongs to class $ j \in [k] $. Then, releasing the histogram of the data is equivalent to calculating $\sum_{i=1}^n h_j(D_i) $ for all $ j \in [k] $.\footnote{Actually, we only need to calculate $\sum_{i=1}^n h_j(d_i)$ for $j \in [k-1]$ since by the definition of a histogram we have $\sum_{j=1}^k h_j(d) = 1$ for all $d \in \cD$.}

\modified{Let $\cQ$ denote the set of all product distributions for $X$, i.e., $\mathcal Q \coloneqq \{P_{X} \colon P_{X} = \prod_{i=1}^n P_{D_i} \}$, where $P_{D_i}$ denotes the marginal distribution of the $i$-th entry.} Note that if $X \sim P_X \in \cQ$, then its entries are independent but not necessarily identically distributed. We assume that the probability of each class is bounded \modified{away from zero}. Specifically, let $\alpha \in (0, \frac{1}{k}]$ be a constant, and assume $P_X \in \mathcal{Q}_\alpha$, where
\begin{multline*}
    \mathcal{Q}_\alpha = \Big\{P_X \in \cQ : P_{D_i} (\{d \in \mathcal D : h_j(d) = 1\}) \geq \alpha, \\
    \text{ for all } i\in [n], j \in [k]  \Big\}.
\end{multline*}
The value of $\alpha$ directly affects our privacy bound, with larger values yielding stronger guarantees.
\begin{theorem}
\label{thm:histo}
Suppose $X = (D_1, \ldots, D_n)$ is a database drawn according to $P_X \in \cQ_\alpha$. Let $Y^k = (Y_1, \ldots, Y_k)$ be the outcome of the Laplace mechanism with scale parameter $b >0$ releasing the histogram of the data, i.e., 
\begin{equation*}
    Y_j = \sum_{i=1}^n h_j(D_i) +N_j, \quad j \in [k],
\end{equation*}
where $N_j \sim \mathrm{Lap}(b)$. Then, the PML of each record in the database is bounded by 
\begin{equation}
\label{eq:pml_hist}
    \ell_{\mathrm{Hist}}(D_i \to y^k) \leq \frac{2}{b} - \log \big(1 - \alpha + \alpha \exp\big(\frac{2}{b} \big) \big),
\end{equation}
for all $i \in [n]$, distributions $P_X \in \mathcal Q_\alpha$, and $y^k \in \mathbb R^k$.  
\end{theorem}

\begin{IEEEproof}
\modified{
Let $y^k \in \bR^k$ be the outcome of the mechanism. Without loss of generality, we calculate the information leaking about the first entry in the database $D_1$. Let   
\begin{equation*}
    S_j \coloneqq y_j - \sum\limits_{i=2}^n h_j(D_i), \quad j \in [k]. 
\end{equation*}
The definition of $S_j$ allows us to separate the effect of $D_1$ on each $y_j$ from all the other entries. Furthermore, let 
\begin{equation*}
    p_j \coloneqq P_{D_1} (\{d \in \cD : h_j(d) = 1 \}), \quad j \in [k],
\end{equation*}
denote the probability that $D_1$ belongs to class $j$. 

Now, fix some $d_1 \in \cD$ belonging to class 1, so that $h_1(d_1) = 1$ and $h_j(d_1) = 0$ for all $2 \leq j \leq k$. Using the definition of the Laplace mechanism and the fact that the database entries are independent, we write 
\begin{align*}
    &\frac{M_{Y \mid D_1}(y^k \mid d_1)}{M_Y(y^k)}= \frac{\mathbb E\Big[\exp \Big(- \frac{\big\lvert S_1 - 1 \big\rvert}{b}\Big) \prod\limits_{j=2}^k \exp \Big(- \frac{\big\lvert S_j \big\rvert}{b} \Big) \Big]}{\mathbb E\Big[\prod_{j=1}^k  \exp \Big(- \frac{1}{b} \big\lvert S_j- h_j(D_1) \big\rvert\Big)\Big]}\\
    &= \frac{\mathbb E\Big[\exp \Big(- \frac{1}{b} \big\lvert S_1 - 1\big\rvert \Big) \prod\limits_{j=2}^k \exp \Big(- \frac{1}{b} \big\lvert S_j \big\rvert \Big) \Big]}{\bE \Big[ \mathbb E\Big[\prod_{j=1}^k  \exp \Big(- \frac{1}{b} \Big\lvert S_j - h_j(D_1) \Big\rvert\Big) \;\Big\vert \; D_1 \Big] \Big]}\\
    &= \frac{\mathbb E\Big[\exp \Big(- \frac{1}{b} \big\lvert S_1 - 1 \big\rvert \Big) \prod\limits_{j=2}^k \exp \Big(- \frac{1}{b} \big\lvert S_j\big\rvert \Big)\Big]}{\sum_{t=1}^k p_t \cdot \bE\Big[\exp \Big(- \frac{1}{b} \lvert S_t - 1 \rvert \Big) \prod\limits_{j \neq t}  \exp \Big(- \frac{1}{b} \lvert S_j \rvert \Big)\Big]}\\
    &= \Bigg(p_1 + \sum_{t=2}^k p_t \cdot \\
    &\quad \frac{\bE\Big[\exp \!\Big(\!\!- \frac{\lvert S_1 \rvert}{b} \Big) \exp \!\Big(\!\!- \frac{\lvert S_t - 1 \rvert}{b} \Big) \!\!\!\prod\limits_{j \in [k] \setminus \{1,t\}}  \!\!\!\exp \!\Big(\!\!- \frac{\lvert S_j \rvert}{b} \Big)\Big]}{\mathbb E\Big[\exp \!\Big(\!\!- \frac{\lvert S_1 - 1\rvert}{b}\Big) \exp \!\Big(\!\!- \frac{\lvert S_t \rvert}{b} \Big) \!\!\!\prod\limits_{j \in [k] \setminus \{1,t\}} \!\!\!\exp \!\big(\!\!- \frac{\lvert S_j \rvert}{b}  \Big)\Big]} \Bigg)^{-1}\\
    &\labelrel \leq{eq:proof_1} \Big(p_1 + \sum_{t=2}^k p_t \exp\big(-\frac{2}{b}\big)\Big)^{-1}\\
    &= \exp \left(\frac{2}{b} - \log \; \Big(p_1 \cdot \exp\big(\frac{2}{b} \big) + 1-p_1 \Big) \right)\\
    &\labelrel \leq{eq:proof_2} \exp \left(\frac{2}{b} - \log \; \Big( \alpha \cdot \exp\big(\frac{2}{b} \big) + 1 - \alpha \Big) \right),
\end{align*}
where \eqref{eq:proof_1} follows by applying the triangle inequality to
\begin{gather*}
    \abs{S_1} \leq \lvert S_1 - 1\rvert + 1, \quad \text{and} \quad \lvert S_t - 1 \rvert \leq \abs{S_t} + 1, 
\end{gather*}
and \eqref{eq:proof_2} is due to the fact that the mapping $p_1 \mapsto - \log \Big(p_1 \cdot \exp(\frac{2}{b} )+ 1-p_1\Big)$ is deceasing in $p_1$, and $p_1 \geq \alpha$. 

Finally, noticing that the above argument can be replicated for $d_1$ belonging to any class, we get the bound 
\begin{align*}
    \ell_{\mathrm{Hist}}(D_1 \to y^k) &= \log \; \max_{d_1} \; \frac{M_{Y \mid D_1}(y^k \mid d_1)}{M_Y(y^k)}\\
    &\leq\frac{2}{b} - \log \; \big( 1 - \alpha + \alpha \exp \big(\frac{2}{b} \big) \big).
\end{align*}}
\end{IEEEproof}
\modified{Importantly, the bound in Theorem~\ref{thm:histo} is tight in the sense that if the distribution of $X$ satisfies $P_{D_i} (\{d \in \mathcal D : h_j(d) = 1\}) = \alpha$ for some $i \in [n]$, then there exists an outcome $y^k$ such that the upper bound on $\ell_{\mathrm{Hist}}(D_i \to y^k)$ is attained.} 

Using inequalities $e^x \geq 1 + x$ and $\log (1+x) \geq x - \frac{x^2}{2}$ for $x \geq 0$ we can further upper bound \eqref{eq:pml_hist} to get
\begin{equation}
\label{eq:pml_hist_simple}
    \ell_{\mathrm{Hist}}(D_i \to y^k) \leq \frac{2 (1-\alpha)}{b} + \frac{2 \alpha^2}{b^2},
\end{equation}
for all $i \in [n]$, $P_X \in \mathcal Q_\alpha$, and $y^k \in \mathbb R^k$. \modified{Observe that $\frac{2}{b}$ is the DP parameter of the Laplace mechanism for queries with sensitivity $2$. This matches the bounds in~\eqref{eq:pml_hist} and~\eqref{eq:pml_hist_simple} as $\alpha \to 0$, i.e., when the database entries follow an arbitrary product distribution (see also~\cite[Thm.~4.2]{inferential_privacy}). Hence, Theorem~\ref{thm:histo} shows that restricting $P_X$ to $\cQ_\alpha$ reduces the information leakage roughly by a factor of $1 - \alpha$.}

\modified{
\subsection{Empirical Analysis}
\label{sec:empirical}
\begin{figure}[t]
    \centering
    \begin{subfigure}[b]{0.5\textwidth}
        \includegraphics[width=0.85\textwidth]{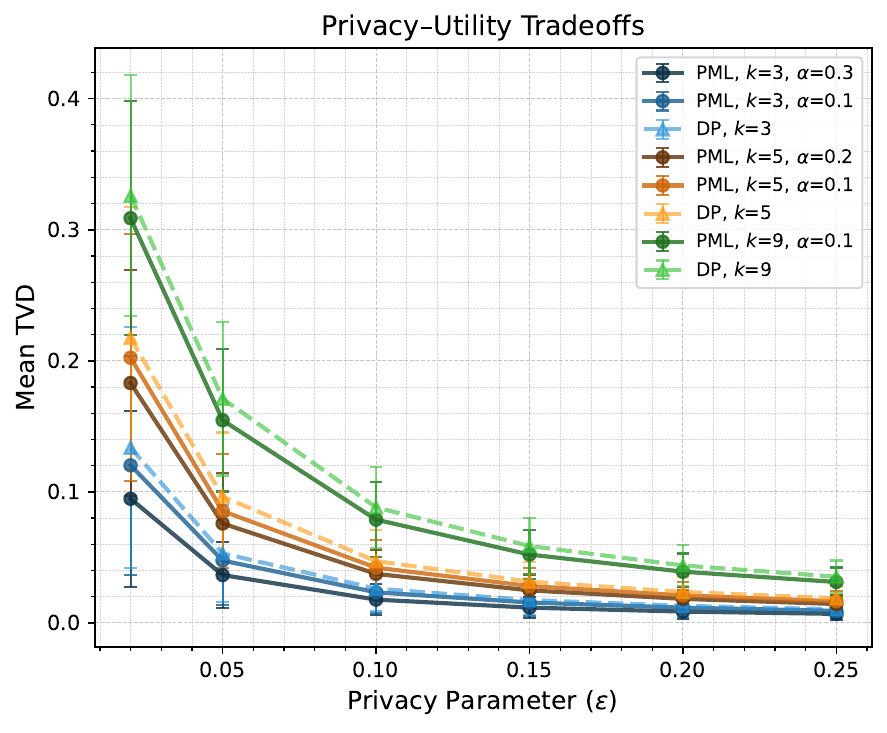}
        \caption{TVD vs $\varepsilon$}
        \label{fig:subfig1}
    \end{subfigure}
    \hfill
    \begin{subfigure}[b]{0.5\textwidth}
        \includegraphics[width=0.85\textwidth]{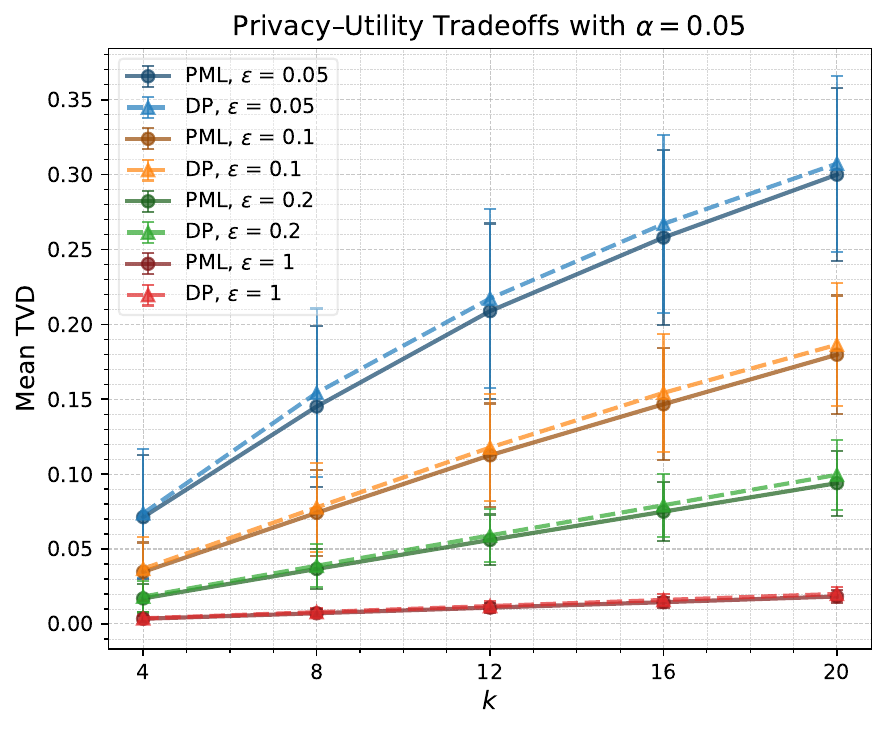}
        \caption{TVD vs $k$}
        \label{fig:subfig2}
    \end{subfigure}

    \caption{Comparison of PML-privatized and DP-privatized histograms: (a) TVD as a function of privacy level $\varepsilon$; (b) TVD as a function of histogram dimensionality $k$.}
    \label{fig:empirical}
\end{figure}

Here, we empirically compare the utility of PML-privatized and DP-privatized histograms. We construct a database of $n = 1000$ entries drawn uniformly from $k$ classes. In each experiment, Laplace noise scales $b, b' > 0$ are chosen so that $\varepsilon_{\mathrm{DP}} (b) = \varepsilon_{\mathrm{PML}}(b')  = \varepsilon$ for some $\varepsilon > 0$, where $\varepsilon_{\mathrm{DP}} (b) = \frac{2}{b}$, and 
\begin{equation}
    \varepsilon_{\mathrm{PML}} (b) = \frac{2}{b} - \log \big(1 - \alpha + \alpha \exp\big(\frac{2}{b} \big) \big). 
\end{equation}
Privatized histograms are then clipped to non-negative values and rounded to the nearest integer.

We report two sets of experiments. In the first, we compute the \emph{total variation distance} (TVD) between the privatized and true empirical distributions while sweeping over $\varepsilon$ for multiple values of $k$ and $\alpha$. In the second, we fix $\alpha = 0.05$ and sweep over $k$ for different values of $\varepsilon$. Each experiment is repeated 10,000 times, and we report the mean TVD along with error bars in Figure~\ref{fig:empirical}. It can be observed that in all experiments, PML consistently achieves lower TVD than DP for the same privacy level. The utility gap becomes particularly pronounced in the high-privacy regime, i.e., when $\varepsilon$ is very small. Hence, our results position PML as a compelling choice for achieving better utility under strong privacy guarantees. Moreover, as seen in Figure~\ref{fig:subfig1}, increasing $\alpha$ further reduces the TVD incurred by the PML-adapted mechanism. This is because stronger assumptions about the data distribution enable more effective noise calibration and, consequently, improved utility.}

\section{Discussion and Conclusions}
\label{sec:conclusion}
We draw two important insights from our results. First, observe that~\eqref{eq:pml_hist} is decreasing in $\alpha$, implying that distributions with more balanced probabilities enjoy stronger privacy guarantees for a fixed $b$ \modified{(alternatively, higher utility for fixed privacy level as illustrated in Figure~\ref{fig:subfig1})}. This phenomenon is a recurring theme in the PML framework and has been noted in other contexts, such as the \emph{randomized response} mechanism \cite{kairouz2016extremal,grosse2024extremal} and the \emph{truncated geometric} mechanism~\cite{ghosh2009universally,saeidian2023pointwise_it}. Theorem~\ref{thm:histo} precisely quantifies the gap between the PML and DP parameters and suggests that when assumptions about the data-generating distribution are available, they should be incorporated into the privacy analysis and design to achieve more favorable privacy-utility tradeoffs.

\modified{Second, Theorem~\ref{thm:histo} highlights the advantage of performing a direct privacy analysis rather than relying on composition bounds. A histogram with $k$ bins can be treated as $k - 1$ independent counts. A simplistic approach would then apply the PML composition theorem~\cite[Lemma 1]{saeidian2023pointwise_it} to the counting query~\cite[Prop. 4.6]{inferential_privacy}, yielding
\begin{equation*}
    \ell_{\mathrm{Hist}}(D_i \to y^k) \leq (k - 1) \cdot \left(\frac{1 - \alpha}{b} + \frac{\alpha^2}{2 b^2} \right),
\end{equation*}
for all $i \in [n]$, $P_X \in \mathcal{Q}_\alpha$, and $y^k \in \mathbb{R}^k$. This bound can be significantly looser than the one in~\eqref{eq:pml_hist_simple}. A comparable situation arises in DP: applying composition to the counting query yields a privacy level of $\frac{k - 1}{b}$, which is much larger than $\frac{2}{b}$. These issues of overly conservative composition bounds have also been studied in the context of user-level DP, where adaptive contribution control has been proposed to reduce cumulative privacy loss~\cite{rameshwar2024improving}.}

Finally, in practice, the $1 - \alpha$ improvement in privacy is effective only when $k$ is small. Regardless, adding noise directly to the histogram bins is only viable when $k$ is small. When $k$ is large, the counts can become excessively noisy and uninformative. In such cases, clustering-based approaches or noise addition in transformed domains are preferable. Analyzing these methods within the PML framework is an interesting direction for future work.

\bibliographystyle{IEEEtranN}
\footnotesize
\balance
\bibliography{main}

\end{document}